\documentstyle[preprint,tighten,eqsecnum,aps]{revtex}

\newcommand{\tselea}[1]{\label{#1} \\ & \mbox{ } &
\mbox{(#1)} \nonumber}
\newcommand{\tseleq}[1]{\mbox{ (#1) } \label{#1}}

\newcommand{\tseref}[1]{(\dr{#1})} 
\newcommand{\tsecite}[1]{\cite{#1}} 

\newcommand{\tsebibitem}[1]{\bibitem{#1}} 

\newcommand{\href}[2]{{#2}{}}

\newcommand{\tsecom}[1]{}
\newcommand{\tsenote}[1]{}

\newcommand{\be}{\begin{equation}}
\newcommand{\ee}{\end{equation}}
\newcommand{\ba}{\begin{eqnarray}}
\newcommand{\ea}{\end{eqnarray}}
\newcommand{\dle}[1]{\label{#1}}
\newcommand{\dla}[1]{\label{#1}}
\newcommand{\dr}[1]{\ref{#1}}
\newcommand{\dc}[1]{\cite{#1}}

\newtheorem{lemma}{Lemma}
\newtheorem{theo}[lemma]{Relation}

\newcommand{\half}{{\frac{1}{2}}}

\newcommand{\psizp}{\psi^{(0+)}}
\newcommand{\psizm}{\psi^{(0-)}}
\newcommand{\Nz}{N^{(0)}}

\begin{document}

\draft

\typeout{--- Title page start ---}

\preprint{\parbox{6cm}{Imperial/TP/97-98/16 \\ DAMTP-97-149 \\ hep-ph/9801404}}

\title{Wick's Theorem for
non-symmetric normal ordered products and
contractions}

\author{
T.S. Evans$^a$\thanks{E-mail: \href{mailto:T.Evans@ic.ac.uk}{{\tt T.Evans@ic.ac.uk}}, 
WWW: \href{http://euclid.tp.ph.ic.ac.uk/links/time}{{\tt http://euclid.tp.ph.ic.ac.uk/$\sim$time}}},
T.W.B.Kibble$^a$\thanks{E-mail: {\tt t.kibble@ic.ac.uk}} and 
D.A.Steer$^{b}$\thanks{E-mail: {\tt D.A.Steer@damtp.cam.ac.uk}} }

\address{
\href{http://euclid.tp.ph.ic.ac.uk}{a) Theoretical Physics}, 
Blackett Laboratory, Imperial College, Prince Consort Road,\\ 
London, SW7 2BZ, 
U.K.}

\address{b) D.A.M.T.P., Silver Street, Cambridge, CB3 9EW, U.K.}

\date{\today} 

\maketitle 

\begin{abstract}

We consider arbitrary splits of field operators
into two parts; $\psi = \psi^+ + \psi^-$, and use
the corresponding definition of normal ordering introduced in \dc{ES}. 
In this case the normal ordered products and contractions have none of
the special symmetry properties assumed in existing proofs of
Wick's theorem.  Despite this, we
prove that Wick's theorem still holds in its usual form as long as the
contraction is a
$c$-number.  Wick's theorem is thus shown to be much more general than
existing derivations suggest, and we discuss possible simplifying
applications of this result.

\end{abstract} 


\typeout{--- Main Text Start ---}

\section{Introduction}

\tsenote{Say that in the PI approaches, never worry about Wick at
all}Wick's theorem is at the core of perturbative calculations in the
canonical operator approach to quantum field theory:  on
expansion of the $S$-matrix, it is used to relate time ordered products
of field operators to normal ordered products and contractions
\dc{Wi}.  Within the traditional definition of normal ordered
products, the interaction picture field
operators, ${\psi}(x)$, are split into positive $\psizp$ and
negative $\psizm$
energy waves containing annihilation and creation operators
respectively;
\be
{\psi}(x) = {\psizp}(x) + {\psizm}(x) .
\dle{splitgen}
\ee
Normal ordering, $\Nz [\bullet]$,\tsenote{Say why have the superscript
zero?} then places the
annihilation (creation) operators to the right (left), otherwise
leaving the order of the operators unchanged.  Hence the vacuum
expectation values of normal ordered products are guaranteed to vanish,
\be
\langle 0 | \Nz [\bullet] | 0 \rangle = 0.
\ee
Calculations therefore proceed in a straightforward manner
by simply taking the vacuum expectation value of the operator identity
called Wick's theorem.  

In the literature, Wick's theorem has been proved either through
the use of a generating functional \dc{LeB,IZ,NO}, or by the manipulation
of products of operators \dc{Wi,AGD,FW,Otherprematsubara}.  In both
cases, an essential ingredient of the proof is the fact that the
normal ordered products defined above are {\em symmetric} in the
following sense.  Let us
denote the normal ordered product of $n$ (possibly different) field
operators $\psi_i := {\psi}(x_i)$ evaluated at different space
time points $x_i$ by
\be
\Nz_{1,2,\ldots,n} := \Nz \left[ {\psi}_1 {\psi}_2 \ldots {\psi}_n \right].
\ee
Then since
$\left[ {\psizp}(x), {\psizp}(x') \right]_{\sigma} =
\left[ {\psizm}(x), {\psizm}(x') \right]_{\sigma} = 0$ $\forall x,
x'$ (where $\sigma = +1$ for bosonic operators, and $\sigma = -1$
for fermionic ones),\tsenote{can we make this
statement more general?} it follows that
\be
\Nz_{1,2,\ldots,n} = (-1)^p \Nz_{a,b,\ldots,t} \; \; \; \; \forall n>0.
\dle{symmcon}
\ee
Here $a,b,\ldots,t$ is any permutation of $1,2,\ldots,n$ and $p$ is the
number of times pairs of fermionic fields are interchanged in moving 
from an order $1,2,\ldots,n$  to an $a,b,\ldots,t$ order of fields. 
Normal ordered products satisfying (\dr{symmcon}) are said to be
{\em symmetric}.\tsenote{Put in reason why this is so important in the
derivation of Wick?}  The reason why this symmetry is
central to the existing derivations of Wick's theorem will be
explained in detail in section 3 below where we work with the generating
functional approach.  However, the following comments may help
illustrate the reason.  Recall that Wick's theorem relates time
ordered products, normal ordered products and contractions.  One way
to prove this theorem \dc{Wi,FW} is to consider initially a given order of
times.  The resulting (single) contribution from the time ordered
product may then shown to be equivalent to a combination of normal
ordered products and contractions.  {\em If} these normal ordered products
and contractions are symmetric, then each contribution to the time
ordered product takes the same form, and thus Wick's theorem is proved
\dc{Wi,FW}.  However, if they are {\em not symmetric} this is no 
longer the case (see section 3).


\tsenote{This is what was here before: It occurs when finishing the
derivation of the generating functional, namely in moving from 
equation \tseref{wp6} to
\tseref{wp8}, and then the real problem occurs when using the generating
functional to 
extract the usual form of Wick's theorem i.e.\ in moving from \tseref{p7} to
\tseref{wickns}. }

In general, however, normal ordering with respect to annihilation and
creation operators does {\em not} simplify calculations.  As an
example, consider how, in the presence of symmetry breaking,
one must normal order with respect to new annihilation and creation
operators related by a Bogoliubov
transformation to the ones appearing in the field \dc{AGD,FW}.  This is
equivalent to choosing to split the field in a different way from 
\tseref{splitgen}.
A more recent example \dc{ES} was provided by
a system in thermal equilibrium at a temperature
$T$.  There the thermal expectation value of a normal ordered
product of creation and annihilation operators ($\hat{a}^{\dagger}$
and $\hat{a}$ respectively) no longer vanishes in the way that it
did at $T=0$:
\be
\ll N^{(0)}[\hat{a},\hat{a}^{\dagger}] \gg \; = n_{be}, n_{fd} \neq 0.
\ee
(Here $n_{be}$, $n_{fd}$ are the Bose-Einstein and Fermi-Dirac
distributions respectively.)  As a result, perturbative
calculations at finite $T$ would not seem to be simplified by
the use of Wick's theorem.  This
point has caused some confusion in the literature \dc{FW,Ma,Ga} and it was
addressed in \dc{ES}.  There it was shown that Wick's theorem
could be used in the usual way at finite temperature provided one
split the field in a different way from the usual one into 
creation and annihilation
operators (\dr{splitgen}).  

Thus in general one must consider arbitrary splits of the field
operators.  The precise split used will reflect the physical
environment of the problem if it is to simplify practical
calculations in the usual way.  We denote the arbitrary split by
\be
{\psi}(x) = {\psi}^+(x) + {\psi}^-(x),
\dle{splitgen2}
\ee
where ${\psi}^+(x)$ and ${\psi}^-(x)$ are unspecified in this paper,
so that we can accommodate any
physical problem.  We still call the two
parts `positive' and `negative' even though
the split is general and no longer necessarily into positive and
negative energy waves.  The corresponding
(natural) generalisation of normal ordering is then expressed in
terms of 
${\psi}^+(x)$ and ${\psi}^-(x)$, and is defined to be 
\begin{itemize}
\item {\bf Normal ordering} of products of fields, $N[\bullet]$, 
places all the `positive' parts to the
right and the `negative' parts to the left, otherwise leaving the order
of the fields unchanged.  A sign change occurs whenever two fermion
field operators are interchanged.
\end{itemize}
Note that this reduces to the usual definition, $\Nz[\bullet]$, 
when the split is
into the annihilation and creation operators appearing in the field.
 We have used the $(0)$
superscript to denote this traditional and very special case. 
So, for example, in the
two-point case one has 
\begin{equation} N \left[ \psi_1 \psi_2 \right] := N_{1,2} =
\psi_{1}^{+} \psi_{2}^{+} +
\psi_{1}^{-} \psi_{2}^{+} + \sigma \psi_{2}^{-} \psi_{1}^{+} +
\psi_{1}^{-} \psi_{2}^{-}.
\dle{N2}
\end{equation}  
Such a generalisation of normal ordering was successfully used for
thermal field theory in \tsecite{ES}, and is equivalent to 
discussions of normal ordering involving Bogoliubov transformations and
symmetry breaking \dc{AGD,FW}.\tsenote{Or indeed in the Thermo Field Dynamics approach to
thermal field theory.}
%
%
Contractions are then defined in the usual way through the two-point time
ordered products which, being independent of the split, 
take the canonical form;
\be
T_{1,2} := T \left[ \psi_1 \psi_2 \right]   = \psi_1
\psi_2 \theta(t_1 - t_2) + \sigma \psi_2 \psi_1 \theta(t_2 - t_1).
\tseleq{T2}
\ee
(In general the
$\theta$ functions may be defined for complex times and hence be
contour dependent -- as for example in Euclidean quantum field theory 
and thermal field theory \dc{LeB,LvW}.  Thus the time ordered products
are also contour dependent.  However, the analysis presented here
holds regardless of contour.)
Using (\dr{N2}) and (\dr{T2}) the contraction $D_{1,2}$ is:
\ba
D_{1,2} & =  & D\left[\psi_1,\psi_2\right] := T_{1,2} - N_{1,2}
\dla{W2}
\\ & = & \theta(t_1 - t_2) \left[ \psi_{1}^{+}, \psi_{2}^{-} \right]
_{\sigma}
\nonumber
\\ &  & - \; \theta(t_2 - t_1) \left\{
\left[ \psi_{1}^{+}, \psi_{2}^{+} \right] _{\sigma} +
\left[ \psi_{1}^{-}, \psi_{2}^{+} \right] _{\sigma} +
\left[ \psi_{1}^{-}, \psi_{2}^{-} \right] _{\sigma} \right\}.
\dla{W2b}
\ea
Now the key point is that for such arbitrary splits, the contraction and
two-point normal
product are {\em not} generally symmetric:  from
(\dr{N2}) and (\dr{W2b}) the condition for
symmetry is that for all times $t_1,t_2$
\be
D_{1,2} = \sigma D_{2,1} ~ ~ , ~ ~ N_{1,2} = \sigma
N_{2,1}
\Longleftrightarrow 
\left[ \psi_{1}^{+}, \psi_{2}^{+} \right] _{\sigma} +
\left[ \psi_{1}^{-}, \psi_{2}^{-} \right] _{\sigma} = 0, 
\dle{Nsymm}
\ee
which is not necessarily satisfied.  If these two-point relations
are true, then all higher order normal ordered products are
symmetric in the sense of 
(\dr{symmcon}).  This is only easy to prove if one has already proved that
Wick's theorem holds independent of the symmetry of all normal
ordered products!   

It is conceivable that in some situations, for example
systems out of equilibrium, calculations may be simplified by the
use of a split for which the normal ordered products are not
symmetric ({\em i.e.}\ (\dr{Nsymm}) is not satisfied).  Indeed, out
of equilibrium it may be possible to find a split such that (\dr{Nsymm})
is true for specific times, but without time-translation symmetry
it is doubtful that this split will ensure (\dr{Nsymm}) will be true for all times.
Wick's theorem, which relates normal ordered
products (\dr{N2}), time ordered products (\dr{T2}), and
contractions (\dr{W2b}), has only been proved for the special case
where the normal ordered products and contractions are symmetric.
The question this paper asks then is: Does the
usual form of Wick's theorem still hold when normal ordered
products are not symmetric?
  
Our answer is yes.  Wick's
theorem is in fact more general than the existing proofs
suggest, namely:  {\em if the contraction is a
$c$-number (and hence the split is linear in creation and
annihilation operators), Wick's theorem holds for both symmetric and
non-symmetric normal ordered products.}  Possible practical applications
of this result, in particular to time dependent 
non-equilibrium systems, are discussed in the conclusions.

In section 2 we derive the generating functional for Wick's theorem
in terms of the generalised definition of normal ordering.  We
stress the places in which the non-symmetry of products means that our
derivation differs from others existing in the literature.  In section
3 we use the generating functional to obtain Wick's theorem.  For
symmetric products the proof is immediate -- for non-symmetric ones it
is considerably more complicated.   We finally summarise our work and also comment
on a claim made in the original paper on thermal field theory by
Matsubara \dc{Ma}. Applications for our work are then discussed.

\section{The generating functional for Wick's Theorem}

In this section we derive the generating functional for Wick's
Theorem -- that is, we prove the identity 
\ba
\lefteqn{ T \left[ \exp \{ -i \int d^{4} x \; j(x)
\psi(x) \}
\right]  = 
N \left[ \exp \{ -i \int d^{4} x \; j(x) \psi(x) \} 
\right] }
\nonumber
\\
& \times &
\exp \{ - \frac{1}{2} \int d^{4} x \; d^{4} y \;
j(x) D [ \psi(x) \psi(y) ] j(y) \}.
\dla{p1}
\ea
Here $ \int d^{4} x = \int dt \int d^{3} {\bf x}$, and the precise
range of integration does not affect our 
arguments.  For simplicity, we just consider bosonic fields.
(Comments regarding fermionic fields will be made later.)\tsenote{In
general time integral may be over an arbitrary contour in complex time
plane.}  Thus the $j(x)$'s are $c$-number sources\tsenote{
If the operators are fermionic, the sources are Grassmann variables.
Have to be careful in this case with the order in which you write 
$j \psi \neq \psi j$.}  and since Wick's theorem is used in
perturbation theory, the field operators $\psi(x)$ are in the interaction
picture.  The normal ordered products $N[\bullet]$ and contraction $D
[\psi(x) \psi(y) ]$ have been defined above
 in terms of the arbitrary split of the fields
into positive and negative parts; $\psi(x) = \psi^+(x) +
\psi^-(x)$.  Following the discussion in the previous section, we
therefore make no assumptions about their symmetry properties.  
The first part of the proof presented here follows closely
that of \dc{LeB,IZ}.\tsenote{pp 337 and 416 of Le Bellac, and p179-185 of 
Itzykson and Zuber \tsecite{IZ}.}\tsenote{Le Bellac references - this
was LeB2!}

Recall that for two operators
$A$ and $B$ 
\be
e^{A+B} = e^A e^B e^{\half [B,A]}
\dle{CBH}
\ee
if and only if
\be
[A,[A,B]] = [B,[A,B]] = 0.
\ee
Recall, too, that time ordering puts the larger times to the left, and
so for example in the case of two operators $A(t_1)$ and $A(t_2)$ with
$t_2 > t_1$,
\be
 T\left[ e^{A(t_1) + A(t_2)}\right] = e^{A(t_2)} e^{A(t_1)} =
T\left[ e^{A(t_1)} e^{A(t_2)}\right].
\tseleq{twooporder}
\ee
We comment that a similar relation holds for 
any operation which places the two operators in a given order. 
This will be used for normal ordered products below in equation \tseref{wp5}.

Consider the LHS of (\dr{p1}),
\be
T\left[ e^{-i\int dt  \int d^3 x \; j(x) \psi(x)} \right] := 
T\left[ e^{-i\int dt \; O(t)} \right]
\dle{T1con}.
\ee
Let $t_i$ and $t_f$ denote the initial and final time.  
Divide the total time interval into
$N$ equal intervals with $\Delta t = \frac{t_f - t_i}{N}$ and
\be
t_f = t_N > t_{N-1} > \ldots > t_1 > t_0 = t_i.
\ee
%
%
%
By definition,
\be
\int_{t_i}^{t_f} dt O(t) = \lim_{N \rightarrow \infty} \Delta t
\sum_{j=1}^{N}O(t_j).
\ee
So here,
\ba
T\left[ e^{-i \int_{t_i}^{t_f} dt \;  O(t)} \right] 
& = & 
\lim_{N \rightarrow \infty} T\left[ e^{-i \Delta t 
\sum_{j=1}^{N}O(t_j)} \right]
\nonumber
\\
& = & 
\lim_{N \rightarrow \infty} \left[ e^{-i \Delta t \; O(t_N)} e^{-i
\Delta t\; O(t_{N-1})} \ldots e^{-i \Delta t \;O(t_1)} \right]
\dla{wp2}
\\
& = & 
\lim_{N \rightarrow \infty} \left[  e^{-i \Delta t
\sum_{k=1}^{N}O(t_k)} e^{- \half (\Delta t)^2 \sum_{1<k<l<N}
\left[ O(t_l), O(t_k) \right]} \right]
\dla{wp3}.
\ea
In going from (\dr{wp2}) to (\dr{wp3}) we have used (\dr{CBH}), which is
legitimate since $\left[ O(t_l), O(t_k) \right]$ is a
$c$-number.\tsenote{The reason for this is that the sources are
$c$-numbers, whilst the fields are in the interaction picture and so
can be expanded in terms of annihilation and creation operators.}  
Note also that in this exponential
$t_l > t_k$.  Thus using the
definition of $O(t)$, (\dr{wp3}) gives
\be
T\left[ e^{-i \int d^4 x \;  j(x) \; \psi(x)} \right] =
e^{-i \int d^4 x \;  j(x) \; \psi(x) } 
e^{-\half \int d^4 x \int d^4 y\;  
\theta(x_0 - y_0) j(x) \; \left[ \psi(x) ,
\psi(y) \right] \; j(y)}.
\dle{wp4}
\ee

In order to relate the time ordered product in (\dr{wp4}) to a normal
ordered product, consider an arbitrary split of the fields into `positive' and
`negative' parts introduced in (\dr{splitgen2}) and the corresponding
definition normal ordering given in
the previous section.  Using (\dr{CBH}), the first exponential on the
RHS of (\dr{wp4}) may be re-written as
\ba
\lefteqn{ e^{-i \int d^4 x \;  j(x) \; \psi(x) }  } 
\nonumber
\\
& = &
e^{-i \int d^4 x \;  j(x) \; \left\{ \psi^-(x) + \psi^+(x) \right\} }
\nonumber
\\
& = & 
e^{-i \int d^4 x \;  j(x) \; \psi^-(x)} e^{-i \int d^4 x \;  j(x) \;
\psi^+(x)} 
\nonumber
\\
&& \; \; \; \; \; \; \; \; \; \;  \times
e^{\half \int d^4 x \int d^4 y\;  j(x) \; \left[\psi^-(x),
\psi^+(y)
\right] \; j(y) } 
\nonumber
\\
& = & 
N \left[  e^{-i \int d^4 x \;  j(x) \; \psi(x)} \right] 
e^{\half \int d^4 x \int d^4 y\;  j(x) \; \left[\psi^-(x),
\psi^+(y)
\right] \; j(y)}.
\dla{wp5}
\ea
In obtaining (\dr{wp5}) we have used (\dr{twooporder}) for normal
ordered products, and (\dr{CBH}), which requires that
\be
\left[ \psi^{\pm},\left[ \psi^+,\psi^-\right]\right] = 0.
\ee
This condition is satisfied if $\psi^+$ and $\psi^-$ are linear in
annihilation and creation operators, and in turn it ensures that the
contraction is a $c$-number.  
Substituting (\dr{wp5}) into (\dr{wp4}) gives
\ba
\lefteqn{ T \left[ e^{-i \int d^4 x \;  j(x) \; \psi(x)} \right]  = 
N \left[  e^{-i \int d^4 x \;  j(x) \; \psi(x)} \right] }
\nonumber
\\
&  \times &
e^{-\half \int d^4 x \int d^4 y\; j(x) \; \left\{
\theta(x_0 - y_0)  \left[ \psi(x) ,
\psi(y) \right] \; - \left[\psi^-(x), \psi^+(y)
\right]\right\} j(y)}.
\dla{wp6}
\ea
Our proof now diverges from that given in the usual texts
\dc{LeB,IZ} since we are considering the possibility of having an {\em
arbitrary} (unsymmetric) split of the fields rather than a (symmetric)
one into annihilation and creation operators.  Look at the part in
curly brackets in the second exponential of (\dr{wp6}); it is
\ba
\lefteqn{\theta(x_0 - y_0) 
\left[
\psi(x) ,
\psi(y) \right] \; - \left[\psi^-(x), \psi^+(y)
\right]}
\nonumber
\\
& = &
\left[\psi^+(y) ,\psi^-(x)\right] +  \theta(x_0 - y_0) \left\{ 
\left[ \psi^+(x) , \psi^+(y) \right] + \left[ \psi^-(x) ,
\psi^+(y) \right]
\right.
\nonumber
\\
&&  \; \; \; + \left. 
\left[ \psi^+(x) , \psi^-(y) \right] + \left[ \psi^-(x) ,
\psi^-(y)
\right]
\right\}
\nonumber
\\
& = &
\theta(y_0 - x_0) \left[\psi^+(y) ,\psi^-(x)\right] 
-\theta(x_0 - y_0) \left\{ 
\left[ \psi^+(y) , \psi^+(x) \right] \right.
\nonumber
\\
&&  \; \; \; 
+  \left. \left[ \psi^-(y) , \psi^+(x) \right] + 
\left[ \psi^-(y) , \psi^-(x) \right] \right\}
\nonumber
\\
& = & D\left[  \psi(y) \psi(x) \right]
\dla{wp7},
\ea
where we have used the definition of the contraction in (\dr{W2b}).

Thus, now substitute (\dr{wp7}) into the final exponential of
(\dr{wp6}), relabel $x \leftrightarrow y$ and use the fact that the
$j(x), j(y)$ are $c$-numbers to switch their positions\tsenote{SO what
about fermions?  Have we said we prove this only for bosons?}.  This
enables (\dr{wp6}) to be re-written as 
\ba
\lefteqn{ T \left[ \exp \{ -i \int d^{4} x \; j(x)
\psi(x) \}
\right]  = 
N \left[ \exp \{ -i \int d^{4} x \; j(x) \psi(x) \} 
\right] }
\nonumber
\\
& \times &
\exp \{ - \frac{1}{2} \int d^{4} x \; d^{4} y \;
j(x) D [ \psi(x) \psi(y) ] j(y) \},
\dla{wp8}
\ea
thus proving (\dr{p1}).  This equation is also true for fermionic
operators.  There the sources are Grassmann variables, and the
contraction, defined in (\dr{W2b}), contains anti-commutators.

\tsecom{CUT???: (TSE Don't understand this)   In standard texts, the split
considered is always into annihilation and creation operators.  As a result the
contraction is just a zero temperature Green's function which is symmetric in its
indices.  In that case, when considering for example a scalar field and its
hermitian conjugate, the symmetry can be used to remove the factor of
$1/2$ multiplying the contraction above as one can drop one of the two
equal terms.  Since we are considering general splits
for which the contraction is {\em not} symmetric, that factor of $1/2$ cannot be
removed.}

\section{Wick's theorem for non-symmetric
products}

Wick's theorem is obtained by functionally differentiating (\dr{wp8})
with respect to the sources $j(x)$.  Observe that we need only work with
an even number of fields as the perturbation is about the
expectation value of the fields, and hence the expectation value 
of an odd number of fields will always vanish. 

\tsecom{IS THIS BIT STILL TRUE IN GENERAL -- I.E.\ WHEN NOT IN THERMAL
EQUILIBRIUM?  HOW DO WE AREGUE TO JUST USE AN EVEN NUMBER OF
INDICES???  Recall first that Green functions are thermal expectation
values of products of fields.  We are working in the interaction
picture and so with free fields: hence the thermal expectation value
of an odd number of fields always vanishes.  Thus we need only
consider in the following products involving an even number of
fields. }

Functionally differentiate (\dr{wp8}) with respect to the sources $2n$
times and then set every $j(x_a) = j_a$ to zero.
Since time-ordered products of fields
evaluated at the same time points are undefined, the indices $a$ refer
to different space-time points and are therefore all different.  The LHS of
(\dr{wp8}) gives
\ba
T \left[ \left. \frac{\delta }{\delta j_1 \ldots \delta
j_{2n}}  \exp \{ -i j \psi \} \right]
\right|_{j_a = 0}
& =&  
T \left[ \frac{\delta }{\delta j_1 \ldots \delta
j_{2n}} \frac{(-i j \psi)^{2n}}{(2n)!} \right]
\dla{p5a} 
\\
& = &
 \sum_{\rm perm\, \{a\}} (-1)^p \frac{T_{a_1,a_2,\ldots,a_{2n}}}{(2n)!}
\dla{p5}
\\
& = &
T_{1,2,3,\ldots,2n}
\dla{p5bis}.
\ea
In (\dr{p5}) the sum takes
$a_1,
\ldots a_{2n}$ through all permutations of
$1, \ldots ,2n$;
\be
{\rm perm} \, \{a\} = \protect\left( \begin{array}{ccccc}
 1 & 2 &  3 & \ldots & 2n \\
 a_1 & a_2 & a_3 & \ldots & a_{2n} \\
\end{array} \protect\right)
\ee
and $p$ is the number of times one has to interchange
pairs of fermions in moving from $1, \ldots ,2n$ to $a_1, \ldots
a_{2n}$ order.  The last identity (going from (\dr{p5}) to
(\dr{p5bis})) follows from the fact that by definition time ordered
products are symmetric in their indices, and that there are $(2n)!$
permutations of the numbers
$1, 2,
\ldots 2n$.  This type of manipulation will {\em not} hold for 
normal ordered products and contractions since they are {\em not}, in
general, symmetric in their indices.

Carrying out a similar operation for the RHS of (\dr{wp8}) gives
a general form of Wick's theorem for non-symmetric products;
\be
T_{1,2, \ldots , 2n} = \sum_{m=0}^{n} \sum_{\rm perm\, \{a\}} (-1)^p 
\left[ \frac{N_{a_1,a_2,\ldots, a_{2m}}}{(2m)!} .
\frac{1}{(n-m)!}
\left( \prod_{j=m+1}^{n} \frac{1}{2}D_{a_{2j-1},a_{2j} }
\right) \right] .
\dle{p7}
\ee

In the traditional split of the fields into annihilation and creation
operators, the normal ordered products 
$N^{(0)}_{1,2,\ldots}$
and contraction $D^{(0)}_{1,2}$ are {\em symmetric} as was discussed
in section 1.  As a result, the sum over permutations on the RHS of (\dr{p7})
can easily be evaluated and thus (\dr{p7}) rapidly reduces
to the well known form of Wick's theorem.  For two bosonic fields this
is
\be
T_{1,2} = N^{(0)}_{1,2} + D^{(0)}_{1,2},
\dle{p8}
\ee
and for four bosonic fields
\ba
T_{1,2,3,4}  & = & N^{(0)}_{1,2,3,4} + D^{(0)}_{1,2}N^{(0)}_{3,4} +
D^{(0)}_{1,3}N^{(0)}_{2,4} +D^{(0)}_{1,4}N^{(0)}_{2,3} + D^{(0)}_{2,3}N^{(0)}_{1,4} +
\nonumber
\\
& &
D^{(0)}_{2,4}N^{(0)}_{1,3} + D^{(0)}_{3,4}N^{(0)}_{1,2} + D^{(0)}_{1,2}D^{(0)}_{3,4} +
D^{(0)}_{1,3}D^{(0)}_{2,4} + D^{(0)}_{1,4}D^{(0)}_{2,3} .
\dla{p9}
\ea
Higher order relations follow the same pattern.  Note that in the above
expressions we have {\em chosen} to arrange the indices on any one
given normal ordered product and contraction such that the smaller
indices are to the left of the larger ones.  In the case of symmetric
products, this ordering is of course not important\tsenote{In the
original proof of the operator form of Wick's theorem
\dc{Wi} and in all subsequent proofs (see for example
\dc{LeB,IZ,NO}),\tsenote{Here again some LeB2 reference was used. 
Is this his book or something else?????} symmetric products have always been assumed.}. 
However, it will also be used below where, starting from (\dr{p7}), we
show that the above expressions (equations (\dr{p8}) and (\dr{p9})) hold even for {\em non-symmetric}
products
$N$ and
$D$, {\em as long as the contraction is a c-number}.  We again
consider bosonic fields.  The results below can be extended to
fermionic fields by inserting a minus sign every time the positions of
two fermionic fields are interchanged.  The following relations and
lemmas are useful for the proof.

\subsection{Lemmas and relations}

\begin{lemma}
Two point normal ordered products and contractions are related by
\be
N_{2,1} - N_{1,2}  =
D_{1,2} - D_{2,1}
\dle{p11}.
\ee
\label{lemma1}
\end{lemma}
{\bf Proof:}  By definition of the contraction (\dr{W2})
\ba
D_{1,2} - D_{2,1} & = &\theta_{1,2} \left[ \psi_{1}^{+}, \psi_{2}^{-}
\right]
 - \; \theta_{2,1} \left\{
\left[ \psi_{1}^{+}, \psi_{2}^{+} \right]  +
\left[ \psi_{1}^{-}, \psi_{2}^{+} \right]  +
\left[ \psi_{1}^{-}, \psi_{2}^{-} \right]  \right\}
\nonumber
\\
& -&  \theta_{2,1}  \left[ \psi_{2}^{+}, \psi_{1}^{-} \right]
 + \; \theta_{1,2} \left\{
\left[ \psi_{2}^{+}, \psi_{1}^{+} \right]  +
\left[ \psi_{2}^{-}, \psi_{1}^{+} \right]  +
\left[ \psi_{2}^{-}, \psi_{1}^{-} \right]  \right\}
\nonumber
\\
& =&  \left[\psi_{2}^{+} , \psi_{1}^{+} \right] 
+ \left[\psi_{2}^{-} , \psi_{1}^{-} \right]
\nonumber
\\
& = & N_{2,1} - N_{1,2},
\dla{p11b}
\ea
where $\theta_{1,2} = \theta(t_1-t_2)$ and where the last line
follows directly from the definition of normal ordering (\dr{N2}).

\begin{lemma}
Let $X$ and $Y$ be arbitrary operators. Then if the contraction is a $c$-number,
\be
N\left[X \psi_{1} \psi_{2} Y \right]
= N \left[X \psi_{2} \psi_{1} Y \right] + (D_{2,1} -
D_{1,2})N\left[XY\right].
\dle{p12}
\ee
\label{lemma2}
\end{lemma}
{\bf Proof:}  By definition
\be
N\left[X \psi_{1}^+ \psi_{2}^- Y \right] = N\left[X
\psi_{2}^- \psi_{1}^+  Y \right] \; \; \; \; \forall X,Y.
\dle{p13}
\ee
Therefore
\ba
N\left[X \psi_{1} \psi_{2} Y \right] & = &
N\left[X \psi_{1}^+ \psi_{2}^+ Y \right] + N\left[X \psi_{1}^+
\psi_{2}^- Y \right] 
\nonumber
\\
& & + N\left[X \psi_{1}^- \psi_{2}^+ Y \right] + N\left[X \psi_{1}^-
\psi_{2}^- Y \right]
\nonumber
\\
& = & 
N\left[X \psi_{2}^+ \psi_{1}^+ Y \right] + N\left[X \psi_{2}^-
\psi_{1}^+ Y \right]
\nonumber
\\
& & + N\left[X \psi_{2}^+ \psi_{1}^- Y \right] + N\left[X \psi_{2}^-
\psi_{1}^- Y \right]
\nonumber
\\
&& + N\left[X \left[\psi_{1}^+, \psi_{2}^+\right] Y \right]
+ N\left[X \left[\psi_{1}^-, \psi_{2}^-\right] Y \right]
\nonumber
\\
& = &
N\left[X \psi_{2} \psi_{1} Y \right] + (D_{2,1} -
D_{1,2})N\left[XY\right],
\dla{p14}
\ea
{\em provided} $\left[\psi_{1}^+, \psi_{2}^+\right]$ and
$\left[\psi_{1}^-, \psi_{2}^-\right]$ are $c$-numbers.  This implies that both
$\psi^+$ and $\psi^-$ are linear in the creation and annihilation operators, and
hence that {\em the contraction is a
$c$-number}.
\\
\begin{theo} \label{relation1}
The number of ways of filling $p$ indistinguishable bags each with exactly two
marbles labelled $1,2,\ldots,2q$ is precisely
\be
[(2q)(2q-1)][(2q-2)(2q-3)] \ldots [(2q-2p+2)(2q-2p+1)]
\frac{1}{p!} \frac{1}{2^p}
=
\frac{(2q)!}{(2q-2p)! 2^p p!}
\dle{p15}.
\ee
\end{theo}
{\bf Consequence 3.1}  
 Let labels $a_i$'s take values from
$1, 2, 3, \ldots 2q$.
The number of distinct configurations of $p$ contractions
$D_{a_1,a_2}D_{a_3,a_4}$$ \ldots D_{a_{2p-1},a_{2p}}$ with unordered indices
is given by (\dr{p15}).
\tsenote{The number of distinct configurations with ordered indices is
\be
\frac{(2q)!}{(2q-2p)!p!}.
\ee}
\\
\\
{\bf Consequence 3.2} 
 For a product of $p$ contractions and one normal ordered product,
with a total number of $2q$ indices;
$N_{a_1,a_2,\ldots,a_{2q-2p}} D_{a_{2q-2p+1},a_{2q-2p+2}} 
\ldots D_{a_{2q-1},a_{2q}}$, the number of distinct configurations with
unordered indices is again given by (\dr{p15}).
\tsenote{The number of distinct configurations with ordered indices is
\be
\frac{(2q)!}{(2q-2p)!p!}\times (2q-2p)! = \frac{(2q)!}{p!}.
\ee}
\\
\\
Using these, we now show that (\dr{p8}), (\dr{p9}) and the usual higher order
version of Wick's theorem holds even for non-symmetric normal ordered products and
contractions, as long as the contraction is a $c$-number.
\subsection{Proof for $n=1$}

This is the easiest case.  For $n=1$, (\dr{p7}) gives
\be
T_{1,2} = \frac{1}{2} ( N_{1,2} + N_{2,1}  +  D_{1,2} +  D_{2,1}).
\dle{p10}
\ee
Substituting (\dr{p11}) into (\dr{p10}) gives
\be
T_{1,2} = N_{1,2} + D_{1,2}
\dle{p11bis}
\ee
showing that for $n=1$ (\dr{p8}) holds even for non-symmetric normal ordered
products and contractions.  (This result is of course consistent with
the definition of the contraction;  from (\dr{W2}), $D_{1,2} := T_{1,2} -
N_{1,2}$ and similarly $D_{2,1} := T_{2,1} - N_{2,1} = T_{1,2} - N_{2,1}$. 
Adding these two equations gives (\dr{p10}).)

\subsection{Proof for $n=2$}

For $n=2$, (\dr{p7}) gives
\ba
T_{1,2,3,4} &= &\sum_{\rm perm\, \{a \}}  \sum_{m=0}^{2}
\left[ \frac{N_{a_1,a_2,\ldots, a_{2m}}}{(2m)!} .
\frac{1}{(2-m)!}
\left( \prod_{j=m+1}^{2} \frac{1}{2}D_{a_{2j-1},a_{2j} }
\right) \right]
\nonumber
\\
& = &
\sum_{\rm perm\, \{a \}} \left( \frac{1}{4!} 
N_{a_1,a_2,a_3,a_4 }
 +
\frac{1}{2!}\frac{1}{2} N_{a_1,a_2}D_{a_3,a_4 } +
\frac{1}{2!}\frac{1}{2}\frac{1}{2}D_{a_1,a_2}D_{a_3,a_4 } \right)
\nonumber
\\
& = &\sum_{\rm perm\, \{a \}}
\frac{1}{4!}\left(  N_{a_1,a_2,a_3,a_4 } + 6 N_{a_1,a_2}D_{a_3,a_4 }
+ 3 D_{a_1,a_2}D_{a_3,a_4 } \right)
\dla{p17}
\\
& = & \sum_{\rm perm\, \{a \}} \frac{1}{4!} S_{a_1,a_2,a_3,a_4 }
\dla{p18}
\ea
where in the first term of (\dr{p17}) $m=2$, in the second $m=1$,
in the third $m=0$, and $S_{a_1,a_2,a_3,a_4 }$ will be defined later.
Provided we choose $S_{a_1,a_2,a_3,a_4 }$ correctly (see
below), proof that
$T_{1,2,3,4}$ is given by the usual expression ({\em i.e.}\ (\dr{p9}))
even for non-symmetric $N$ and $D$ will then follow directly  from
(\dr{p18}) if we can show that
$S_{a_1,a_2,a_3,a_4 }$ is symmetric under interchange of any two
indices.

To do that, consider initially the second term of (\dr{p17}); $\sum_{\rm
perm\, \{a \}}6 N_{a_1,a_2}D_{a_3,a_4 }$.  The first important point to
notice is that because of the sum over permutations, the order in which
the $a_i$ indices are written is irrelevant; for example
\be
\sum_{\rm perm \, \{a \}}
N_{a_1,a_2} D_{a_3,a_4} = \sum_{\rm perm \, \{a \}} N_{a_1,a_3}D_{a_2,a_4}.
\dle{p19}
\ee
Secondly, the factor of
$6$ can be understood in the following way.  From consequence
3.2 we know that for a product of one normal ordered
product and one contraction with a total of $2q=4$ indices, there are
$4!/2!2 = 6$ distinct configurations with unordered indices.  Thus we
can re-write the second term of (\dr{p17}) as a sum over the 6 distinct
configurations with unordered indices:
\ba
\sum_{\rm perm \, \{a \}}
6 N_{a_1,a_2} D_{a_3,a_4} & = & \sum_{\rm perm \, \{a \}} \left[
D_{a_1,a_2}N_{a_3,a_4} + D_{a_1,a_3}N_{a_2,a_4} +D_{a_1,a_4}N_{a_2,a_3}
\right.
\nonumber
\\
&  & \; \; \; \; \; \; \; \left. +
D_{a_2,a_3}N_{a_1,a_4} +
D_{a_2,a_4}N_{a_1,a_3} + D_{a_3,a_4}N_{a_1,a_2} \right].
\dla{unord4}
\ea
Note that since the indices on the contractions and normal ordered products
are unordered, there was still some freedom in writing down the above
expression.  We have made the particular choice to write the $a_i$ indices
on each individual $N$ and each individual $D$ with the smallest value of
$i$ to the left.\tsenote{There was an error here before -- it said the
smalles value of $a_i$ to the left.  To see that that's wrong, think
about the case for which $a_1=2$, $a_2=1$, and the rest are unchanged.
The first term in (\dr{unord4}) should be $D_{2,1}N_{3,4}$.  If you
were to order with the smallest value of $a_1$ to the left, that would
give the first term as $D_{1,2}N_{3,2}$ which is {\em not} what we want.}
The reason for this will become clear later.

Now carry out a similar procedure for the third term of (\dr{p17}) involving the
product of 2 contractions:  the 3 terms arise from the 3 distinct configurations
of two contractions with unordered indices (see concequence
3.1), and we then chose to write the indices on each
individual contraction such that the smallest value of
$i$ is to the left.  Thus $S_{a_1,a_2,a_3,a_4 }$ in (\dr{p18}) is defined by
\ba
S_{a_1,a_2,a_3,a_4 } & := & N_{a_1,a_2,a_3,a_4} +
D_{a_1,a_2}N_{a_3,a_4} + D_{a_1,a_3}N_{a_2,a_4}
+D_{a_1,a_4}N_{a_2,a_3} + 
\nonumber
\\
& &
D_{a_2,a_3}N_{a_1,a_4} + D_{a_2,a_4}N_{a_1,a_3} +
D_{a_3,a_4}N_{a_1,a_2} 
\nonumber
\\
& &
+ D_{a_1,a_2}D_{a_3,a_4} +
D_{a_1,a_3}D_{a_2,a_4} + D_{a_1,4}D_{a_2,a_3}
\dla{p20}.
\ea
Hence we see that $S_{a_1,a_2,a_3,a_4 }$ is given by the
RHS of (\dr{p9}) where instead of the symmetric normal ordered
products and contractions one uses the more general non-symmetric
normal ordered products and contractions:  this was the reason why (\dr{p17})
was decomposed as above.

To show that $S_{a_1,a_2,a_3,a_4 }$ is symmetric under the interchange of any
two indices, consider a particular order of the indices, say $a,b,c,d$ and then
the effect of $a \leftrightarrow b$ say.  Now
\ba
S_{a,b,c,d} & = & N_{a,b,c,d} + D_{a,b}N_{c,d} + D_{c,d}(N_{a,b}+
D_{a,b})
\nonumber
\\
&& + \; [ D_{a,c}N_{b,d} + D_{b,c}N_{a,d}] + [D_{a,d}N_{b,c} +
D_{b,d}N_{a,c}]
\nonumber
\\
&& + \; [D_{a,c}D_{b,d} + D_{a,d}D_{b,c}]
\dla{p21}.
\ea
The pairs of terms in square brackets are
invariant under interchange
$a \leftrightarrow b$.  Now use (\dr{p11}) and (\dr{p12}) to show that
the first line of (\dr{p21}) is also invariant under $a \leftrightarrow b$:  
\ba
\lefteqn{N_{a,b,c,d} + D_{a,b}N_{c,d} + D_{c,d}(N_{a,b}+
D_{a,b})}
\nonumber
\\
 & = &
N_{b,a,c,d} +( D_{b,a} - D_{a,b} ) N_{c,d} 
\nonumber
\\
&& + D_{a,b}N_{c,d} +
D_{c,d}(N_{b,a}+ D_{b,a})
\nonumber
\\
& = & 
N_{b,a,c,d} + D_{b,a}N_{c,d} + D_{c,d}(N_{b,a}+
D_{b,a})
\dla{p22}.
\ea
Thus $S_{a,b,c,d} = S_{b,a,c,d}$.  One can prove more generally that
$S_{a,b,c,d}$ is invariant under the interchange of any two 
adjacent indices and hence under any of the $4!$ permutations of the
indices.  Thus from (\dr{p19}) and (\dr{p21}) it follows that
\ba
T_{1,2,3,4}  & = & N_{1,2,3,4} + D_{1,2}N_{3,4} +
D_{1,3}N_{2,4} +D_{1,4}N_{2,3} + D_{2,3}N_{1,4} +
\nonumber
\\
& &
D_{2,4}N_{1,3} + D_{3,4}N_{1,2} + D_{1,2}D_{3,4} +
D_{1,3}D_{2,4} + D_{1,4}D_{2,3} .
\dla{p23}
\ea
 Hence, both at the 2- and 4- point level, the
usual form of Wick's theorem holds even if the contraction and normal
ordered product are not symmetric.\tsenote{To use the generating functional in
the 1st place the contraction must be a $c$-number.}

\subsection{Proof for arbitrary $n$}

 At the $n$th order, the proof is a generalisation of that given above
for $n=2$.  However, before proceeding, we use some of the expressions
encountered in the previous subsection to motivate the introduction of two new
pieces of notation.  

The first enables us to write the terms appearing in square brackets on
the RHS of (\dr{unord4}) in a much more compact way.  Recall that in this
square bracket (which we now denote by $Q$), the $i$ subscript on {\em each}
normal ordered product and on each contraction is arranged in increasing order
going from left to right.  The simplest case is when
$a_1 = 1$, $a_2 = 2$, $a_3 = 3$ and $a_4 = 4$, then $Q$ is given by
\be
Q = D_{1,2}N_{3,4} + D_{1,3}N_{2,4} +D_{1,4}N_{2,3}
+
D_{2,3}N_{1,4} +
D_{2,4}N_{1,3} + D_{3,4}N_{1,2}.
\dle{Qdef}
\ee
However, we can re-write this as
\be
Q = \sum_{{\rm perm\, \{b\}} \in B}
D_{b_1,b_2}N_{b_3,b_4}
\dle{bsim}
\ee
where the sum over permutations of $\{b\}$ is {\em restricted} to
include only those permutations which are
an element of the set $B$ defined by
\be
B = \{ {\rm perm}\, \{b\}: \; b_1< b_2,\; b_3 < b_4 \}.
\dle{Bsetdef}
\ee
(It is possible to check that (\dr{bsim}) indeed reduces to
(\dr{Qdef}).)\tsenote{i.e.\ restrict the sum of
permutations to {\em only sum over those permutations of $\{b\}$ for which} $b_1< b_2$ and
$b_3 < b_4$.}  If we now return to the case in which the $a_i$'s are arbitrary
then $Q$ is given by
\be
Q = \sum_{{\rm perm\, \{b\}} \in B}
D_{a_{b_1},a_{b_2}}N_{a_{b_3},a_{b_4}}
\dle{bsim2}
\ee
where $B$ is defined as in (\dr{Bsetdef}).\tsenote{How does this work?
First do the sum over $b$'s subject to the restriction.  You get
exactly the expression in the square brackets of (\dr{unord4}) as the
$b_i$'s just take values from $1,2,\ldots,2m$.  Then you let the
resulting $a_i$'s be what you want.}

We now introduce the second piece of notation whose aim is to enable one to
re-write expressions such as (\dr{bsim2}) in a slightly simpler way (with
fewer subscripts).  Define
\ba
 N(a_1,a_2,\ldots,a_{2n}) & := & N_{a_1,a_2,\ldots,a_{2n}},
\nonumber
\\
 D(a_1,a_2)  & := & D_{a_1,a_2}.
\dla{newnot}
\ea
As an example of the use of this notation, use (\dr{bsim2}) and (\dr{newnot})
to re-write (\dr{unord4}) as
\be
\sum_{\rm perm \, \{a \}}
6 D(a_1,a_2)\; N(a_3,a_4) = 
\sum_{\rm perm \, \{a \}} \sum_{{\rm perm\, \{b\}} \in B}
D(a_{b_1},a_{b_2})\; N(a_{b_3},a_{b_4}).
\dle{fin4}
\ee

We now continue with the proof of Wick's theorem to $n$th order.
\\
\\
\\
$\bullet$ Step 1.

For a given value of $m$, the RHS of
(\dr{p7}) can be re-written as a sum over distinct configurations
of $N$'s and $D$'s, with the indices on each individual 
$N$ and each individual $D$ arranged in increasing order going from
left to right (compare with (\dr{fin4})).  
To do that, use Lemma \tseref{lemma3}:

\begin{lemma}
The following identity holds
\ba
\lefteqn{ \frac{(2n)!}{(2m)! (n-m)! 2^{n-m}} \sum_{\rm perm\, \{a\}}
N(a_1,a_2,\ldots,a_{2m})\prod_{j=m+1}^{n} D(a_{2j-1},a_{2j}) }
\nonumber
\\
&=& \sum_{\rm perm\, \{a\}} \sum_{{\rm perm\, \{b\}} \in B}
N(a_{b_1},a_{b_2},\ldots,a_{b_{2m}})
\prod_{j=m+1}^{n}D(a_{b_{2j-1}},a_{b_{2j}}) 
\ea
where 
\be
{\rm perm} \, \{b\} = \protect\left( \begin{array}{ccccc}
 1 & 2 &  3 & \ldots & 2n \\
 b_1 & b_2 & b_3 & \ldots & b_{2n} \\
\end{array} \protect\right)
\ee
and the restriction on the sum of permutations of $\{b\}$ is such that 1)
it arranges the indices on each individual normal ordered
product and on each individual contraction in increasing order going from left to
right, and 2) it orders the contractions by size of the leading element so as
to avoid repetition of $D_{1,2}D_{3,4} \ldots$ as $D_{3,4} D_{1,2} \ldots$: 
\be
B  = \{ {\rm perm \, \{b\}}: \protect\left[ \begin{array}{l}
b_1<b_2<\ldots<b_{2m} \\
b_{2m+1}<b_{2m+2} \\
b_{2m+3}<b_{2m+4} \\
\ldots \\
b_{2n-1}<b_{2n} \\
\end{array} \protect\right]; b_{2m+1} < b_{2m+3} < \ldots < b_{2n-1} 
\}.
\ee
\label{lemma3}
\end{lemma}
{\bf Proof:}  From Relation \tseref{relation1}, for a product of $n-m$
contractions and one normal ordered product with a total of $2n$
indices, there are
$(2n)!/(2m)!(n-m)!2^{n-m}$ distinct configurations with unordered indices. 
These distinct configurations are represented by the sum over permutations
of the $b$ labels.  On a given contraction or normal ordered product, the
order of the indices is arbitrary, but above the choice has
been made by the restriction.  
\\
\\
$\bullet$ Step 2.

Using Lemma \tseref{lemma3}, re-write the RHS of (\dr{p7}) as
\ba
\lefteqn{ T_{1,2, \ldots , 2n} }
\nonumber
\\
& = & \frac{1}{(2n)!}\sum_{\rm perm \, \{a\}} 
\left[
\sum_{m=0}^{n} \sum_{{\rm perm\, \{b\}} \in B}
N(a_{b_1},a_{b_2},\ldots,a_{b_{2m}})
\prod_{j=m+1}^{n}D(a_{b_{2j-1}},a_{b_{2j}}) \right]
\nonumber
\\
& =: & 
 \frac{1}{(2n)!}\sum_{\rm perm \, \{a\}}  S(\{a\}).
\dla{tdef}
\ea
This is a generalisation of (\dr{p20}) to arbitrary $n$.
\\
\\
$\bullet$ Step 3.

Use Lemmas \tseref{lemma1} and \tseref{lemma2} to show that 
\begin{lemma}
For any two permutations, $\{a\}$ and $\{a'\}$
\be
S(\{a\}) = S(\{a'\}).
\ee
\label{lemma4}
\end{lemma}
{\bf Proof:}  Consider two permutations related by
 $\{a\} \rightarrow \{a'\}$ with
\ba
 a_i & = & a'_{i+1}
\nonumber
\\
a_{i+1} & = & a'_i
\nonumber
\\
a_j & = & a'_j \; \; \forall j \neq i,i+1.
\ea
The only non-zero contributions to $\Delta := S(\{a\}) - S(\{a'\})$ arise when
{\em both} the indices $a_i$ and $a_{i+1}$ are {\em either} on the same
normal ordered product, or on the {\em same} contraction.  The reason
is that when these indices occur on different normal ordered products
or contractions, the restriction on the sum of permutations of $\{b\}$
orders the terms in exactly the same way in both the $\{a\}$ and
$\{a'\}$ permutation thus making $\Delta$ vanish.  (These terms are
the generalisation of those in square brackets in (\dr{p21}).)

Thus we may write
\ba
\lefteqn{  S(\{a\}) - S(\{a'\}) }
\nonumber
\\
& = & \sum_{m=0}^{n} \sum^{\; \; \; \; \; \; \prime}_{{\rm perm\, \{b\}} \in B}
 \left[ N(a_{b_1},a_{b_2},\ldots,a_{b_{2m}}) -
N(a'_{b_1},a'_{b_2},\ldots,a'_{b_{2m}}) \right]
\nonumber
\\
&&\mbox{} \times
\prod_{j=m+1}^{n}D(a_{b_{2j-1}},a_{b_{2j}})
\dla{iip1eletN}
\\
& + &
\sum_{m=0}^{n} \sum^{\; \; \; \; \; \; \prime \prime}_{{\rm perm\, \{b\}} \in B}
 N(a_{b_1},a_{b_2},\ldots,a_{b_{2m}})
\nonumber
\\
&& \mbox{} \times \left[ D(a_i,a_{i+1}) - D(a'_i,a'_{i+1}) \right]
\prod_{\stackrel{j=m+1}{j \neq k} }^{ n }
 D(a_{b_{2j-1}},a_{b_{2j}}).
\dla{iip1eletD}
\ea
Here the prime on the sum in (\dr{iip1eletN}) means that the sum over
permutations of $\{b\}$ is further restricted by the condition that
$(i,i+1) \in \{b_1,b_2,\ldots,b_{2m}\}$.  (Thus it guarantees that the
indices $a_i$ and $a_{i+1}$ only occur on the $N$ and {\em not} on the
$D$'s.)  In (\dr{iip1eletD}), the double prime means that this sum is
further restricted to have $\{i,i+1\}=\{b_{2k-1},b_{2k}\}$ for some
$k>m$, and this is the reason for which the contribution with $(j=k)$
is excluded in the product. (Hence the double prime on the sum ensures
that the indices $a_i$ and $a_{i+1}$ both occur on a single $D$.)

Noting that when $m=0$ there are no contributions to (\dr{iip1eletN}) (since it
involves no normal ordered products), and using Lemma \tseref{lemma2},
(\dr{iip1eletN}) may be re-written as
\ba
\lefteqn{\sum_{m=0}^{n} \sum^{\; \; \; \; \; \; \prime}_{{\rm perm\, \{b\}} \in B}
 \left[ N(a_{b_1},a_{b_2},\ldots,a_i,a_{i+1},\ldots,a_{b_{2m}}) -
N(a'_{b_1},a'_{b_2},\ldots,a'_i,a'_{i+1},\ldots,a'_{b_{2m}}) \right] }
\nonumber
\\
&& \times
\prod_{j=m+1}^{n}D(a_{b_{2j-1}},a_{b_{2j}}) 
\nonumber
\\
&=&\sum_{m=1}^{n}  \sum_{{\rm perm\, \{c\}} \in B}
N(a_{c_1},a_{c_2},\ldots,a_{c_{2m-2}}) \left[ D(a_{i+1},a_i) -
D(a_i,a_{i+1}) \right]
\prod_{j=m}^{n-1} D(a_{c_{2j-1}},a_{c_{2j}}) 
\nonumber
\\
&=&\sum_{p=0}^{n-1} \sum_{{\rm perm\, \{c\}} \in B}
N(a_{c_1},\ldots,a_{c_{2p}})\left[ D(a_{i+1},a_i) -
D(a_i,a_{i+1}) \right]
\prod_{j=p+1}^{n-1}D(a_{c_{2j-1}},a_{c_{2j}}) 
\dla{eqlN}
\ea
where $p=m-1$, $c_j=b_j$ for $j<i$ and $c_j=b_{j+2}$ for $j \geq i$.
\tsenote{In the 1st line of
(\dr{iip1eletN}), I've written the $a_i$ and $a_{i+1}$ next
to each-other as I think the $b$ permutation means that there can be
nothing in between.  In (\dr{iip1eletD}) I haven't written down
$D({i+1},i)$ as the sum over perms $b$ excludes of this.}  

The second term of equation (\dr{iip1eletD}) vanishes for $m=n$ (as
then it contains no contractions) and thus it can be re-written as
\ba
\lefteqn{ \sum_{m=0}^{n} \sum^{\; \; \; \; \; \; \prime \prime}_{{\rm
perm\, \{b\}} \in B} 
 N(a_{b_1},a_{b_2},\ldots,a_{b_{2m}}) }
\nonumber
\\
&& \times \left[ D(a_i,a_{i+1}) - D(a'_i,a'_{i+1}) \right]
\prod_{\stackrel{j=m+1}{j \neq k} }^{ n }
 D(a_{b_{2j-1}},a_{b_{2j}})
\nonumber
\\
&=&
\sum_{m=0}^{n-1} \sum^{\; \; \; \; \; \; \prime \prime}_{{\rm perm\, \{b\}} \in B}
N(a_{b_1},\ldots,a_{b_{2m}})\left[ D(a_{i},a_{i+1}) -
D(a_{i+1},a_{i}) \right]
\nonumber
\\
&&\; \; \; \; \; \; \; \times 
\prod_{\stackrel{j=m+1}{j \neq k} }^{n}D(a_{b_{2j-1}},a_{b_{2j}}) .
\dla{eqlD}
\ea
%
Recalling the definition of the $c$'s, it is now possible to see that
(\dr{eqlN}) and (\dr{eqlD}) just differ by a sign, and thus from (\dr{iip1eletD}) we have proved that
$\Delta = S(\{a\}) - S(\{a'\}) = 0$.  Hence $\Delta$ vanishes
for {\em any} two permutations. 
\\
\\
$\bullet$ Step 4.

Since $S(\{a\}) = S(\{a'\})$ for any two permutations $\{a\}$ and
$\{a'\}$, then from (\dr{tdef}) one has
\ba
T_{1,2, \ldots , 2n} & = &
 \frac{1}{(2n)!}\sum_{\rm perm \, \{a\}}  S(\{a\}) 
\nonumber
\\
&=& S(\{a\})
\tselea{wickns}
\ea
where $S(\{a\})$ is just given by the usual form of Wick's theorem with the
ordered indices.  This finishes the proof that Wick's theorem holds in
its usual form for non-symmetric normal ordered products and
contractions, {\em provided} the
contraction is a $c$-number.



\section{Conclusions}

In this paper we have worked with a general split of the field
operators into two parts and the corresponding non-symmetric normal
ordered products and contractions.  We have shown that, somewhat
surprisingly, Wick's theorem still holds in its usual form for such
non-symmetric products as long as the contraction is a $c$-number.  As
a result, Wick's theorem -- an important calculational tool in field
theory -- is more general than was originally thought.

We now comment on another interesting idea 
inspired by the discussion of Wick's theorem in the original paper on
thermal field theory by Matsubara \tsecite{Ma}.  There, the definition of the
contraction given for bosonic relativistic phonon field is {\em always
symmetric} for all splits, and consistent with another definition of
normal ordering $N^{(T)}[\bullet]$; 
\ba
N^{(T)}[\psi_1 \psi_2 \ldots \psi_n] &=& N[ T[\psi_1 \psi_2 \ldots \psi_n]
].
\tselea{nomat}
\ea
Note that $N^{(T)}[\bullet]$ is distinct from $N[\bullet]$, 
the definition of normal ordering which was given in section 1 and was 
used through out this
paper:  by definition $N^{(T)}$ is always symmetric in the sense of
\tseref{symmcon} because of the
way it is built out of the time-ordered product (hence our
designation with a $(T)$ superscript$^{\dc{foot}}$).  It therefore 
gives a corresponding symmetric contraction
\ba
D^{(T)}[\psi_1,\psi_2] &=& T_{12} - N^{(T)}_{12}
=
\theta (t_1-t_2) [\psi_1^+,\psi_2^-]_\sigma 
+ \sigma \theta (t_2-t_1) [\psi_2^+,\psi_1^-]_\sigma .
\ea

It is clear that there are in fact many possible splits of the field
which may be used to define normal ordering, and even then the 
definition of $N^{(T)}$ shows that one can think of other definitions of
normal ordering different from the $N$ given in section 1.   Whichever 
definition is used, we
have shown that even if the normal ordered product is not symmetric
and the contraction is a $c$-number,
Wick's theorem can be used so that calculations are not too cumbersome.

In practice the only definite cases we know of have symmetric
normal ordered
products.  Symmetry breaking problems expressed in
terms of Bogoliubov transformations \tsecite{AGD,FW} and equilibrium
thermal field theory, whether using path-ordered methods
\tsecite{ES} or  Thermo Field Dynamics \tsecite{TFD}, all use
our definition of $N$
but have symmetric products.  It is easy to see why  symmetric products
are so common; it is because the most useful normal ordered
product is one which has zero expectation value.  Thus taking the
expectation value of Wick's theorem implies that
the $c$-number 
contraction must then be symmetric (since it is then 
equal to the expectation value of the two-point
time-ordered function).  In turn the operator version of Wick's
theorem now shows that the normal ordered
operator must itself be symmetric.

However, there is at least one situation where we
might envisage the appearance of non-symmetric products, and that is
non-equilibrium field theory.  In this case the enviroment is
changing with time so we should expect the split, and hence the
normal ordered products, also to change in time.  It is then not at
all obvious whether one can ever define a
normal ordered products of fields at arbitrary times 
which have zero physical expectation values.  Thus,
reversing the argument of the previous paragraph, it is not at all
clear that symmetric products will simplify the calculations, and
therefore it is reasonable that non-symmetric products may prove to be a
useful tool.

\section*{Acknowledgements}

We thank Brian Steer for many useful comments and for correcting a
number of mistakes in the original manuscript.  T.S.E.\ thanks  the
\href{http://www.royalsoc.ac.uk/}{Royal Society} 
for their support.  D.A.S. is supported by 
\href{http://www.pparc.ac.uk}{P.P.A.R.C.} through a 
research fellowship and is a member of Girton College, Cambridge.  
This work was supported in part by the
European Commission under the Human Capital and Mobility
programme, contract number CHRX-CT94-0423.

\typeout{--- No new page for bibliography ---}

\end{document}